\documentclass[%
 reprint,
 amsmath,amssymb,
 aps,
 superscriptaddress,
]{revtex4-1}

\usepackage{array}
\usepackage{graphicx}
\usepackage{dcolumn}
\usepackage{bm}
\usepackage[thinlines]{easytable}

\usepackage[usenames,dvipsnames]{color}

\usepackage[normalem]{ulem}

\newcolumntype{P}[1]{>{\centering\arraybackslash}p{#1}}

\begin{document}

\preprint{APS/123-QED}

\title{Post-Newtonian Dynamics in Dense Star Clusters: Binary Black Holes in the LISA Band}

\author{Kyle Kremer}
 \email{kremer@u.northwestern.edu}
\affiliation{
 Department of Physics \& Astronomy, Northwestern University, Evanston, IL 60202, USA\\
 }
\affiliation{
 Center for Interdisciplinary Exploration \& Research in Astrophysics (CIERA), Evanston, IL 60202, USA\\
}

\author{Carl L. Rodriguez}
\affiliation{%
 Pappalardo Fellow; MIT-Kavli Institute for Astrophysics and Space Research, Cambridge, MA 02139, USA \\
}%

\author{Pau Amaro-Seoane}
\affiliation{
Institute of Space Sciences (ICE, CSIC) \& Institut d'Estudis Espacials de Catalunya (IEEC)\\
at Campus UAB, Carrer de Can Magrans s/n 08193 Barcelona, Spain\\
Institute of Applied Mathematics, Academy of Mathematics and Systems Science, CAS, Beijing 100190, China\\
Kavli Institute for Astronomy and Astrophysics, Beijing 100871, China }

\author{Katelyn Breivik}
\affiliation{Canadian Institute for Theoretical Astrophysics, University of Toronto, 60 St. George Street, ON M5S 3H8, Canada\\
 }

\author{Sourav Chatterjee}
\affiliation{
 Center for Interdisciplinary Exploration \& Research in Astrophysics (CIERA), Evanston, IL 60202, USA\\
}
\affiliation{Tata Institute of Fundamental Research, Homi Bhabha Road, Mumbai 400005, India}

\author{Michael L. Katz}
\affiliation{
 Department of Physics \& Astronomy, Northwestern University, Evanston, IL 60202, USA\\
 }
\affiliation{
 Center for Interdisciplinary Exploration \& Research in Astrophysics (CIERA), Evanston, IL 60202, USA\\
}

\author{Shane L. Larson}
\affiliation{
 Department of Physics \& Astronomy, Northwestern University, Evanston, IL 60202, USA\\
 }
\affiliation{
 Center for Interdisciplinary Exploration \& Research in Astrophysics (CIERA), Evanston, IL 60202, USA\\
}

\author{Frederic A. Rasio}
\affiliation{
 Department of Physics \& Astronomy, Northwestern University, Evanston, IL 60202, USA\\
 }
\affiliation{
 Center for Interdisciplinary Exploration \& Research in Astrophysics (CIERA), Evanston, IL 60202, USA\\
}

\author{Johan Samsing}
\affiliation{
 Department of Astrophysical Sciences, Princeton University, Princeton, NJ 08544, USA\\
 }

\author{Claire S. Ye}
\affiliation{
 Department of Physics \& Astronomy, Northwestern University, Evanston, IL 60202, USA\\
 }
\affiliation{
 Center for Interdisciplinary Exploration \& Research in Astrophysics (CIERA), Evanston, IL 60202, USA\\
}

\author{Michael Zevin}
\affiliation{
 Department of Physics \& Astronomy, Northwestern University, Evanston, IL 60202, USA\\
 }
\affiliation{
 Center for Interdisciplinary Exploration \& Research in Astrophysics (CIERA), Evanston, IL 60202, USA\\
}

\date{\today}

\begin{abstract}
The dynamical processing of black holes in the dense cores of globular clusters (GCs), makes them efficient factories for producing binary black holes (BBHs). Here we explore the population of BBHs that form dynamically in GCs and may be observable at mHz frequencies or higher with the future space-based gravitational-wave observatory, LISA. We use our Monte Carlo stellar dynamics code, which includes gravitational radiation reaction effects for all BH encounters. By creating a representative local universe of GCs, we show that up to dozens of these systems may be resolvable by LISA. Approximately one third of these binaries will have measurable eccentricities ($e > 10^{-3}$) in the LISA band and a small number ($\lesssim 5$) may evolve from the LISA band to the LIGO band during the LISA mission.
\end{abstract}

\maketitle

\section{Introduction}
\label{sec:intro}

The groundbreaking detections of gravitational waves (GWs) by LIGO/Virgo \citep{Abbott2016a,Abbott2016b,Abbott2016c,Abbott2016d, Abbott2017} have prompted a multitude of studies aimed at understanding the formation mechanisms for merging binary black holes (BBHs). A variety of formation channels have been proposed for these sources, involving isolated massive binary evolution \citep[e.g.,][]{Dominik2012,Dominik2013,Belczynski2016a,Belczynski2016b}, primordial BHs \citep[e.g.,][]{Bird2016,Sasaki2016}, galactic nuclei \citep[e.g.,][]{O'Leary2009,AntoniniRasio2016,Hoang2018}, secular interactions in hierarchical triple systems \citep[e.g.,][]{AntoniniRasio2016,Antonini2017,Silsbee2017,Liu2017, Hoang2018,Leigh2018,RodriguezAntonini2018}, and dynamical formation in dense star clusters \citep[e.g.,][]{PortegiesZwart2000,Banerjee2010,Rodriguez2015a, Rodriguez2016a,Askar2017,Samsing2018c,Fragione2018, Zevin2018}, which is the focus of this study.

Unlike BBHs formed through isolated binary evolution, which are expected to have circularized orbits with component spins nearly aligned with the binary angular momentum, dynamically-formed BBHs can have measurable eccentricities and spin-misalignments that result directly from their formation process. In the cores of dense star clusters, frequent three- and four-body resonant encounters impart large eccentricities \citep[e.g.,][]{Samsing2014, Samsing2018a, Rodriguez2018b} and random spin orientations \citep[e.g.,][]{Rodriguez2016c} to binaries. Hence, if measurable, spin-misalignments and/or eccentricities (or lack thereof) may serve as fingerprints pointing toward the specific BBH formation channel \citep[e.g.,][]{Breivik2016,Rodriguez2016c,Samsing2018a,D'Orazio2018}.

Most LIGO/Virgo sources should be seen with similar properties: small mass ratios,
relatively large masses ($> 10\,M_{\odot}$), small spins, and approximately
zero eccentricity, regardless of the formation scenario \citep{Amaro-Seoane2016}. Even for the majority of dynamically-formed BBHs, eccentricities acquired at the time of dynamical binary formation are largely erased during the GW-driven inspiral \citep{Samsing2018a,Rodriguez2018a,Rodriguez2018b}. More recently, it has been shown that a small fraction of BBH mergers from GCs ($\sim 5\%$) will occur through GW capture events that may have eccentricities in excess of 0.05 in the LIGO band \citep{Lower2018, Samsing2018,Rodriguez2018a,Rodriguez2018b,Zevin2018}, but these systems do not make up a significant fraction of the total population. On the other hand, for the upcoming space-based interferometer LISA \citep{Amaro-Seoane2013,Amaro-Seoane2017}, which will observe similar BBHs at lower GW frequencies ($10^{-5}-1\,$Hz), residual eccentricities may still be apparent for a much larger fraction (up to $\sim 40\%$; \citep{Samsing2018a}) of resolvable sources. The fact that LISA can measure inspiraling sources that merge in the
LIGO/Virgo domain is an idea that has been put forward in the literature. In
the work of \citep{Amaro-SeoaneSantamaria10} this idea was presented in the
context of massive binaries, and after the discovery that LIGO had observed
BBHs with masses larger than the nominal $10\,M_{\odot}$,
\citep{Sesana2016} and \citep{Nishizawa2017} revisited this idea. However,
not all LIGO/Virgo sources are audible by LISA, as \citep{Chen2017} proved,
since LISA is deaf to \textit{highly} eccentric ($e \gtrsim 0.7$) binaries in the mass range of relevance.

Several recent studies have examined the potential population of, in particular, dynamically-formed binaries in the LISA band. In \citep{Banerjee2017,Banerjee2018,Rastello2019}, post-Newtonian (PN) direct $N$-body simulations were used to explore the population of LISA sources assembled dynamically in open clusters. \citep{Kremer2018c} used Monte Carlo GC models (without PN corrections) to show that up to dozens of LISA sources may be found in the Milky Way GCs (including BBHs as well as other compact binaries containing neutron stars and white dwarfs). Most recently, \citep{Samsing2018a}, \citep{D'Orazio2018}, and \citep{Samsing2018b} used semi-analytic methods, including PN corrections, to show that a large fraction of BBH mergers from GCs will have measurable eccentricities in the LISA band. These recent analyses also demonstrated that PN corrections play an important role in the formation and evolution of BH populations in GCs and thus the inclusion of these corrections is important for a full understanding of dynamically assembled LISA sources. 

Here we present our first full-scale GC models, including PN gravity (radiation reaction), to examine BBHs in the LISA band.  Using our Monte Carlo dynamics code, \texttt{CMC}, we create a representative local universe of GCs (out to a distance of 500 Mpc), and examine the BBH systems that may be resolvable by LISA with sufficiently high signal-to-noise ratios. We also explore the distributions of orbital and component properties of LISA BBHs (taking into account PN effects and all dynamical effects expected within realistic GCs) to predict the number of binaries that will have measurable eccentricities in the LISA band. 

In Section \ref{sec:method}, we describe our technique for modeling GCs and describe the models used in this study. In Section \ref{sec:multiband}, we explore all BBH mergers identified in our models and discuss their evolution through the LISA band. In Section \ref{sec:local_universe}, we describe the weighting scheme implemented to simulate a representative local universe and discuss the number of BBHs expected to be resolved by LISA. We discuss our results and conclude in Section \ref{sec:discussion}.

\section{Modeling globular clusters}
\label{sec:method}

To model GCs, we use our H\'{e}non-style Monte Carlo code, \texttt{CMC} (for a review, see \citep{Henon1971a,Henon1971b,Joshi2000,Joshi2001,Fregeau2003,Pattabiraman2013,Chatterjee2010,Chatterjee2013,Rodriguez2015a}). \texttt{CMC} includes all physics relevant to the long-term evolution of GCs including two-body relaxation, stellar evolution (implemented using updated versions of the \texttt{SSE} and \texttt{BSE} packages; \citep{Hurley2000,Hurley2002}), three-body binary formation, and small-$N$ gravitational encounters (implemented using the \texttt{Fewbody} package; \citep{Fregeau2004,Fregeau2007}).

We account for relativistic effects by including the 2.5 PN term (radiation reaction) in three- and four-body dynamical encounters involving more than one BH (which are integrated using the \texttt{Fewbody} code). See \citep{Rodriguez2018a,Rodriguez2018b,Amaro-Seoane2016,Samsing2018a,D'Orazio2018,Zevin2018} for more detail concerning the role of PN effects in these small-$N$ encounters.

Here, we use a $4\times3\times2$ grid of  24 independent GC models identical to those used in \citep{Rodriguez2018b}. The models span four values in initial particle number ($N=2\times 10^5, 5\times 10^5, 10^6, 2\times 10^6$) and three values in galactocentric distance (2 kpc, 8 kpc, and 20 kpc), with corresponding metallicity values of $Z = 0.01\,Z_{\odot},0.05\,Z_{\odot}$, and $0.25\,Z_{\odot}$, respectively. Additionally, we consider two values for the initial cluster virial radius (1 and 2 pc).

We assume an initial binary fraction of $10\%$ for all models with binary orbital periods drawn from a distribution flat in log and initial eccentricities drawn from a thermal distribution. We use the compact object formation prescriptions of \citep{Fryer2001} and \citep{Belczynski2002}. NS natal kicks are drawn as in \citep{Hobbs2005}. BHs are assumed to form with mass fallback and BH natal kicks are reduced in magnitude according to the fractional mass of fallback material (see \citep{Morscher2015} for further detail).

\section{Binary black holes across frequency bands}
\label{sec:multiband}

As discussed in \citep{Rodriguez2018b}, we identify three dynamical formation channels for BBH mergers: (1) Binaries that merge as isolated binaries after dynamically-mediated ejection from their host cluster (henceforth referred to as the ``Ejected" channel); (2) Binaries still retained in their host clusters that merge  \textit{between} resonant dynamical encounters (henceforth referred to as the ``In-cluster'' channel); and (3) Binaries that merge through gravitational capture \textit{during} resonant encounters (henceforth referred to as the ``GW-capture'' channel). These three merger channels have been identified and explored in several recent studies \citep[e.g.,][]{Rodriguez2018b,Samsing2018a,D'Orazio2018,Zevin2018} and it has been shown that these channels can produce binaries of varying orbital parameters (in particular, varying eccentricities) when passing through the LIGO and LISA bands. Note that a small fraction of mergers ($\sim5\%$) occur through primordial binary evolution at early times ($t \lesssim 100\,$Myr); we exclude these primordial mergers as they are not influenced by dynamical interactions (see \citep{Rodriguez2018b} for further detail). 
Because these binaries preferentially merge very early in the evolution of their host clusters and because LISA can only resolve binaries out to $z\approx0.1$ (which limits LISA to older clusters with ages greater than a few Gyr), these primordial BBHs (which by definition merge in their host clusters before being influenced by dynamical interactions) are likely not accessible to LISA. 

\subsection{GW Strain from Eccentric Binaries}

For an eccentric binary, with eccentricity, $e$, semi-major axis, $a$, and component masses, $M_1$ and $M_2$, the characteristic strain at the $n^{\rm{th}}$ harmonic can be written as \citep[e.g.,][]{Barack2004}:

\begin{equation}
\label{eq:hcn}
h_{c,n}^2 = \frac{1}{(\pi D)^2} \Bigg(\frac{2G}{c^3} \frac{\dot{E}_{n}}{\dot{f}_{n}} \Bigg).
\end{equation}
Here, $D$ is the luminosity distance to the source and $f_{n}$ is the rest-frame GW frequency of the $n^{\rm{th}}$ harmonic given by

\begin{equation}
f_{n} = n f_{\rm{orb}}
\end{equation}where $f_{\rm{orb}}$ is the rest-frame orbital frequency. $f_{n}$ is related to the observed (detector frame) GW frequency, $f_{n,\,z}$, by $f_{n} = f_{n,\,z} (1+z)$.

$\dot{E}_{n}$ is the time derivative of the energy radiated in GWs at (rest-frame) frequency, $f_{n}$, which to lowest order is given by \citep[e.g.,][]{Peters1963}:

\begin{equation}
\label{eq:Edot}
\dot{E}_{n} = \frac{32}{5} \frac{G^{7/3}}{c^5}\Big(2 \pi \,f_{\rm{orb}}\, \mathcal{M}_{c} \Big)^{10/3} \,g(n,e)
\end{equation}
where $\mathcal{M}_{c}$ is the (rest-frame) chirp mass, which is related to observed chirp mass, $\mathcal{M}_{c,\,z}$ by

\begin{equation}
\mathcal{M}_{c} = \mathcal{M}_{c,\,z}(1+z)^{-1}=\frac{(M_1M_2)^{3/5}}{(M_1+M_2)^{1/5}}(1+z) ^{-1}.
\end{equation}
To lowest order, $\dot{f}_{n} = n\,\dot{f}_{\rm{orb}}$, where $\dot{f}_{\rm{orb}}$ is found by combining

\begin{equation}
\frac{da}{dt} = \frac{64}{5}\frac{G^3M_1^2 M_2^2 (M_1+M_2)}{c^5 a^5}F(e)
\end{equation}
\citep[e.g.,][]{Peters1964} with the time derivative of Kepler's third law. In this case,

\begin{equation}
\label{eq:fdot}
\dot{f}_{n} = n \frac{96}{10 \pi} \frac{(G\mathcal{M}_{c})^{5/3}}{c^5} \Big(2 \pi \, f_{\rm{orb}} \Big)^{11/3} \,F(e)
\end{equation}
where $F(e) = [1 + (73/24)e^2 + (37/96)e^4]/(1-e^2)^{7/2}$. Combining Equations \ref{eq:hcn}, \ref{eq:Edot}, and \ref{eq:fdot} we the obtain:

\begin{equation}
h_{c,\,n}^2 = \frac{2}{3\pi^{4/3}}\frac{G^{5/3}}{c^3}\frac{M_{c,\,z}^{5/3}}{D^2}\frac{1}{f_{n,\,z}^{1/3}(1+z)^{2}}\Big( \frac{2}{n} \Big)^{2/3} \frac{g(n,e)}{F(e)}
\end{equation}
where we have written the expression in terms of the detector frame (redshifted) chirp mass and GW frequency, $\mathcal{M}_{c,\,z}$ and $f_{n,\,z}$.

\begin{figure}
\begin{center}
\includegraphics[width=0.5\textwidth]{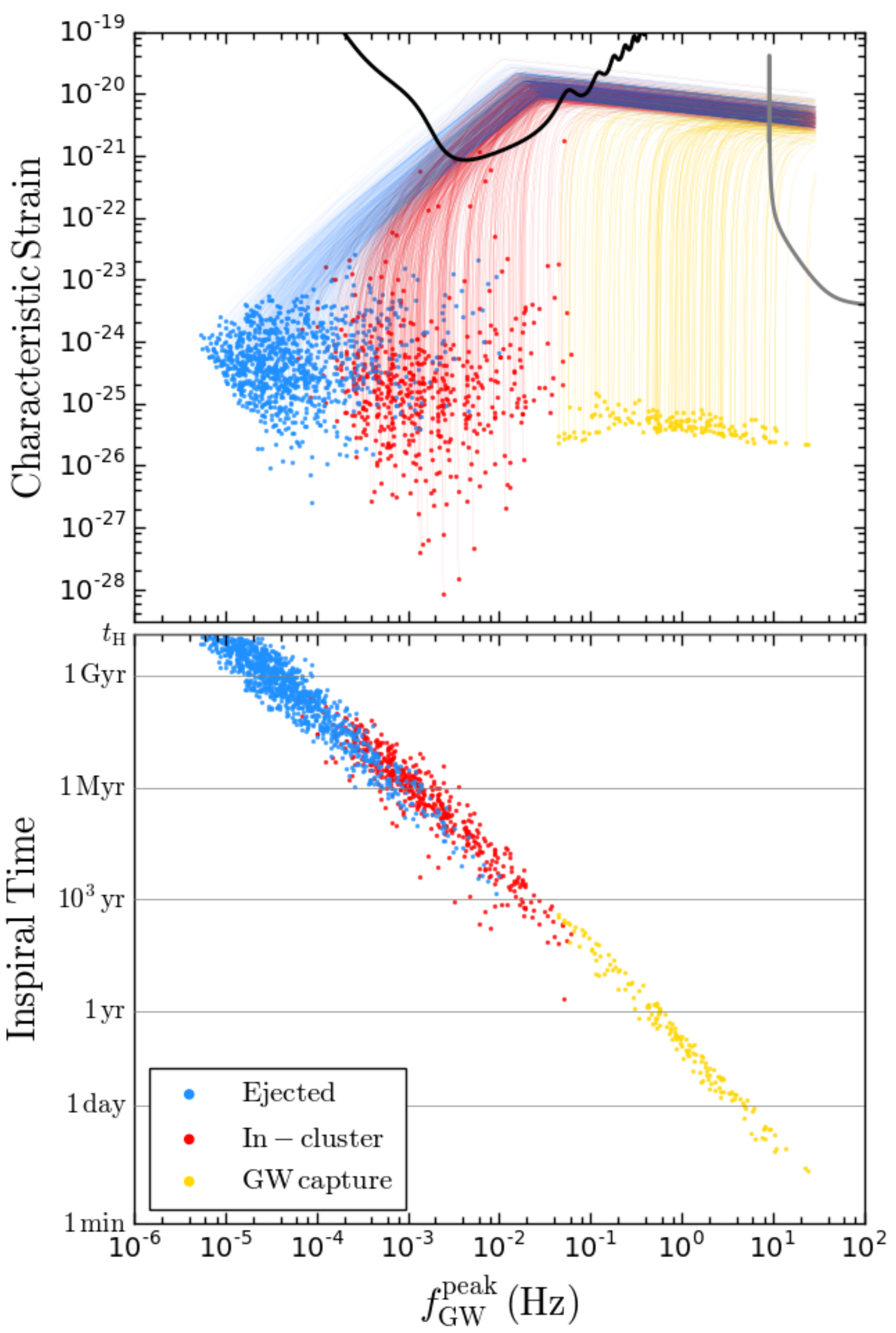}
\caption{\label{fig:multiband} Top panel shows evolution of characteristic strain at frequency of peak emission, $f_{\rm{GW}}^{\rm{peak}}$, for all BBH mergers with $t_{\rm{merge}}<t_{\rm{H}}$ in our models, assuming a distance of 250 Mpc from Earth. Bottom panel shows inspiral time versus $f_{\rm{GW}}^{\rm{peak}}$ at formation. Blue, red, and yellow denote the mergers which occur through the ejected, in-cluster, and GW-capture channels, respectively. Filled circles denote the values at the time of last dynamical encounter. 
The black and gray curves in the top panel denote the LISA and LIGO sensitivity curves, respectively \citep{Cornish2018,Abbott2018}.}
\end{center}
\end{figure}

The characteristic strain, $h_{c,\,n}$, is a measure of the number of cycles per frequency bin of width $\Delta f = 1/T_{\rm{obs}}$, where $T_{\rm{obs}}$ is the LISA mission lifetime (we assume a fiducial value of $T_{\rm{obs}}=5\,$years). For stellar-mass BBHs, the source lifetime can be significantly longer than the observation lifetime, thus it is necessary to take into account the fraction of the mission lifetime a particular source spends within a given frequency bin when discussing the characteristic strain. Therefore, when showing detectability results for particular binaries (as in Figures \ref{fig:multiband} and \ref{fig:LISA}), we show the characteristic strain, $h_{c,\,n}$, multiplied by the square root of $\rm{min}[1,\dot{\it{f}_n}(\it{T}_{\rm{obs}}\,/\,\it{f}_n)\rm{]}$ to account for the frequency band swept by each source during the observation time \citep[see, e.g.,][]{Willems2007,Sesana2005,D'Orazio2018}. 

\subsection{Strain-$f_{\rm{GW}}^{\rm{peak}}$ evolution}

The top panel of Figure \ref{fig:multiband} shows evolutionary tracks for all (dynamical) BBH mergers identified in our models that have merger times less than a Hubble time. These tracks are computed by integrating the equations of purely GW-driven orbital evolution \citep[e.g.,][]{Peters1964} given the binary orbital parameters following the last dynamical encounter, as calculated in \texttt{CMC}.  The tracks in Figure \ref{fig:multiband} show the characteristic strain versus peak frequency of GW emission for eccentric binaries, given by \citep{Wen2003}:

\begin{equation}
\label{eq:f_GW_peak}
f_{\rm{GW}}^{\rm{peak}} = \frac{\sqrt{G(M_1 + M_2)}}{\pi}\frac{(1+e)^{1.1954}}{[a\,(1-e^2)]^{1.5}}.
\end{equation}
Each track shows the evolution from the time of formation (marked by the filled circles) to $f_{\rm{GW}}^{\rm{peak}}=30\,$Hz. The bottom panel of the figure shows the inspiral time for these binaries versus $f_{\rm{GW}}^{\rm{peak}}$ at the time of formation.

The colors in Figure \ref{fig:multiband} denote the three different formation channels, as described in the figure caption. For ejected and in-cluster binaries, the orbital parameters at formation are well-defined: they are simply the binary parameters following the last dynamical encounter. However, for GW capture mergers, the time of binary formation is not well-defined (and if merger occurs through ``direct collision'' of BHs, a binary is never actually formed; see \citep{Rodriguez2018b}), so we simply assume these GW capture binaries form at a reference eccentricity of 0.9999 and integrate these systems backward from the $a$ and $e$ values recorded by the \texttt{fewbody} calculation at a pericenter distance of $100\,M$, where $M$ is the total mass of the BBH. Since the pericenter distance (which determines the peak frequency of GW emission) asymptotically approaches the true (and unknown) initial pericenter distance for these binaries as $e$ approaches 1, the particular choice of the reference eccentricity has no significant effect (see, e.g., \citep{Zevin2018}).

As Figure \ref{fig:multiband} clearly demonstrates, the ejected, in-cluster, and GW capture channels produce BBHs with increasing $f_{\rm{GW}}^{\rm{peak}}$ at the time of formation. In particular, while BBHs produced through the ejected and in-cluster channels form at lower frequencies 
and subsequently evolve through the LISA and then LIGO bands, BBHs creates via GW capture form at higher frequencies 
and typically skip the LISA band entirely. \citep{Chen2017} also noted that some sources detected by ground-based detectors cannot be observed by LISA; for high eccentricity binaries, a detector in the decihertz regime is necessary for co-detection.

Similar tracks in strain-frequency space for these three formation channels were explored in \citep{D'Orazio2018} using semi-analytic methods to model binary-single encounters undergone by BHs in typical GC environments. In contrast, our results arise from realistic, full-scale GC models that span a range in GC properties including metallicity, total mass, virial radius, and galactocentric distance. While \citep{D'Orazio2018} considered only  $30+30 M_{\odot}$ BBHs (as a fiducial case) and a realistic range of BBH formation times and static GC properties, our models produce a self-consistent population of BBHs with a realistic spectrum in BH masses, and their formation times. Satisfyingly, the general trends observed in our strain-frequency diagram agree quite well between our detailed models and their more approximate ones.


\begin{figure}
\begin{center}
\includegraphics[width=0.5\textwidth]{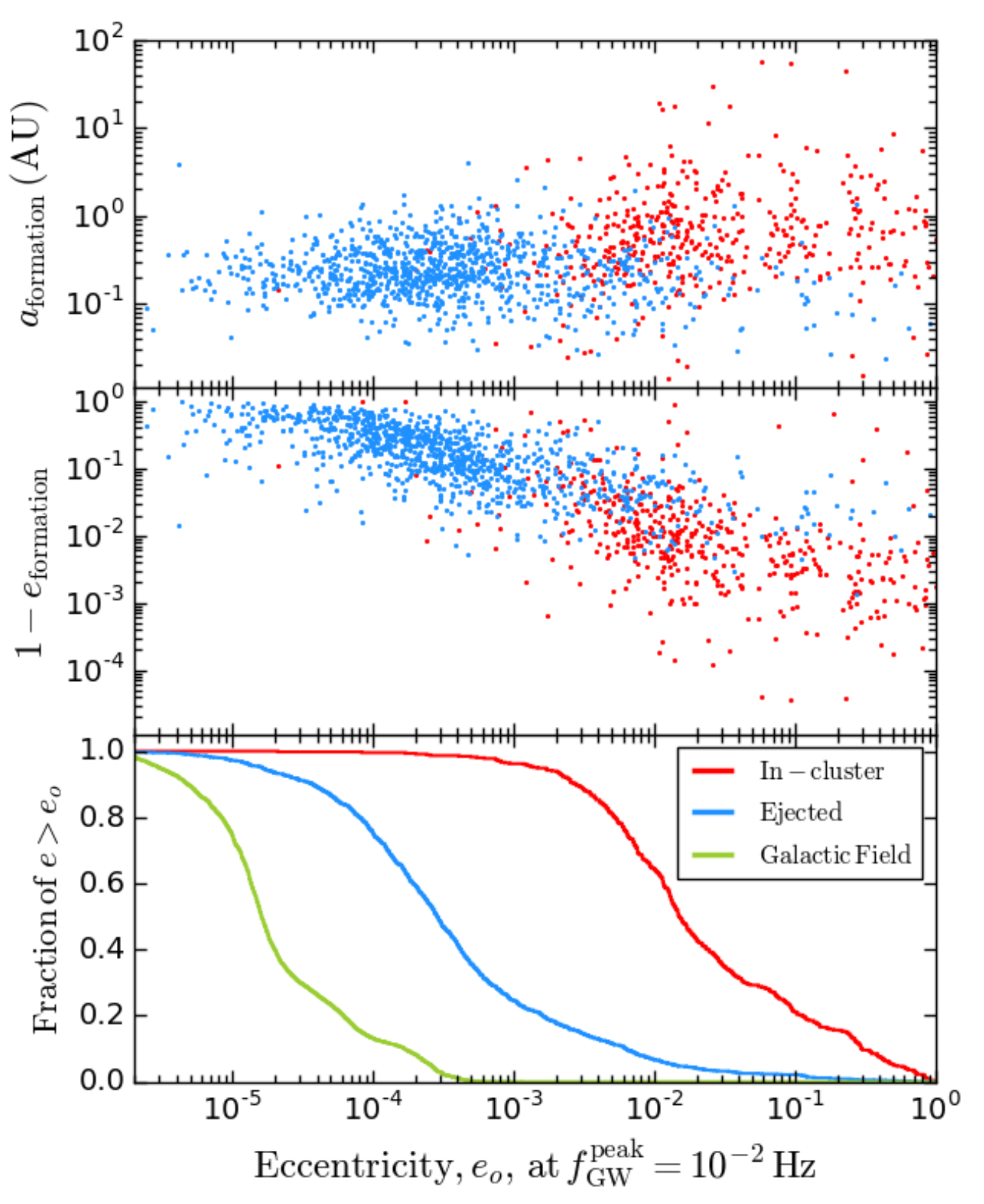}
\caption{\label{fig:3panel} 
The eccentricities of all merging BBHs at a GW frequency of $10^{-2}\,$Hz are shown on the x-axis of all three panels above. As before, red color denotes in-cluster BBH mergers and blue denotes BBHs that merge after ejection from their host cluster. The top panel shows the semi-major axis at the time of formation (defined as the last dynamical encounter) on the y-axis.  The middle panel shows on the y-axis the eccentricity at the time of formation, and the bottom panel shows the cumulative distribution of all sources at $f_{\rm{GW}}^{\rm{peak}}=10^{-2}\,$Hz. The bottom panel also includes the eccentricity distribution for a population of binaries formed in the Galactic field also at $f_{\rm{GW}}^{\rm{peak}}=10^{-2}\,$Hz, created using the \texttt{cosmic} population synthesis code, as described in the text.}
\end{center}
\end{figure}

\subsection{Eccentricity Distribution in LISA}

Figure \ref{fig:3panel} shows the eccentricity distribution for all BBH mergers shown in Figure \ref{fig:multiband} at $f_{\rm{GW}}^{\rm{peak}}=10^{-2}\,$Hz, which approximately corresponds to the so-called ``bucket'' of the LISA sensitivity curve \citep[e.g.,][]{Cornish2018}. The top and middle panels, respectively, show the distributions of semi-major axis and eccentricity at formation (following the last dynamical encounter). All binaries formed through the GW capture channel (yellow systems in Figure \ref{fig:multiband}) form at frequencies in excess of $10^{-2}\,$Hz, so are not shown here.


As the bottom panel of Figure \ref{fig:3panel} shows, $78\% \, $($8\%$) of the in-cluster (ejected) mergers will have $e > 10^{-2}$ at $f_{\rm{GW}}^{\rm{peak}} = 10^{-2}\,$Hz. We also include in the bottom panel the eccentricity distribution at $f_{\rm{GW}}^{\rm{peak}} = 10^{-2}\,$Hz for a population of BBHs formed in the Galactic field. This  population is created using the same binary evolution models as our  GC populations, but is initialized according to a metallicity-dependent Milky Way star formation history based on galaxy {\bf{m12i}} in the Latte simulation suite. The Latte suite of FIRE-2 cosmological, zoom-in, baryonic simulations of Milky Way-mass galaxies \citep{Wetzel2016}, part of the Feedback In Realistic Environments (FIRE) simulation project, were run using the Gizmo gravity plus hydrodynamics code in meshless, finite-mass mode \citep{Hopkins2015} and the FIRE-2 physics model \citep{Hopkins2018}. All binaries in this population, including the binary fraction, are initialized according to observed distributions detailed in \citep{Moe2017}. The population synthesis was performed with \texttt{cosmic} \footnote{https://cosmic-popsynth.github.io/}, a binary population synthesis code designed explicitly to produce realistic Milky Way populations. In particular, \texttt{cosmic} adapts the number of simulated binaries for each binary evolution number such that the shape of the distributions of mass, eccentricity, and orbital period of BBHs at formation converge to a defined shape in the LISA sensitivity band. This is quantified using a convergence criteria, or $match$, defined as
\begin{equation} \label{eq: match}
 match = \frac{\sum\limits_{k=1}^{N} P_{k,i} P_{k,i+1}}{ \sqrt{ \sum\limits_{k=1}^{N} (P_{k,i}P_{k,i})\sum\limits_{k=1}^{N} (P_{k,i+1}P_{k,i+1})}},
\end{equation}
\noindent where $P_{k,i}$ denotes the probability for the $k^{th}$ bin for the $i^{th}$ iteration. As the number of simulated binaries increases, the \textit{match} tends to unity. We continue to simulate binaries until $match>1-10^{-6}$ for the mass, eccentricity, and orbital period distributions which, here, results in a population of $3.15 \times 10^4$ BBHs.

Similar to the results from \citep{Breivik2016}, the eccentricity of BBHs formed in the Galactic field is lower than BBHs formed in GCs that are ejected from or retained in the GC. We plan to perform a more in depth comparison of all observable parameters from these populations, including masses and distance, in a future study.

Note that the Galactic field distribution shown in Figure \ref{fig:3panel} considers only binaries formed through isolated binary evolution and does not consider the contribution of BBH mergers arising from secular evolution of hierarchical triples. Triples may contribute significantly to the overall BBH merger rate and, like the GC channels considered in this analysis, may lead to BBHs with high eccentricities relative to BBHs formed strictly through isolated binary evolution \citep[e.g.,][]{Antonini2017,Randall2018,Fang2019}.

\subsection{Distinguishing Between the Merger Channels}
\label{sec:JS}

As described in Section \ref{sec:multiband}, cluster dynamics produces (at least) three distinct BBH merger channels, which are ejected mergers, in-cluster mergers, and GW capture mergers. As illustrated in Figure \ref{fig:multiband}, each of these channels gives rise to a unique distribution across GW peak frequency space, which can be used to observationally distinguish them from one another. To provide some insight into the relative location of the distributions and how they scale with the cluster properties, we present in this section a few relevant analytical relations to complement our numerical results. We especially focus on how a given channel relates to $f_{\rm{GW}}^{\rm{peak}}$ and $e$ at formation. For this discussion, we use the analytical framework presented in \citep{Samsing2014, Samsing2018d, Samsing2018}. For simplicity, the equal mass limit is assumed and we only include strong encounters up to binary-single.

For deriving $f_{\rm{GW}}^{\rm{peak}}$ and $e$ at the time of formation for a given BBH belonging to a given channel, we start by noticing that each channel is associated with its own characteristic time scale, $\tau$. That is, for an assembled BBH to contribute to a particular channel, its GW merger time, $t_{\rm GW}$ (see Figure \ref{fig:multiband}), has to be comparable to the associated time $\tau$, where $\tau \sim$ the Hubble time ($t_{\rm H}$) for the ejected mergers, $\tau \sim$ the time between strong encounters ($t_{\rm enc}$) for the in-cluster mergers, and $\tau \sim$ the BBH orbital time ($T_{\rm orb}$) for the GW capture mergers \citep[e.g.][]{Samsing2018a, D'Orazio2018}. Given that $f_{\rm GW}^{\rm{peak}} \approx \pi^{-1} \sqrt{{2GM}/{r_{\rm p}^3}}$, where $M$ is the BH mass and $r_{\rm p}$ is the BBH pericenter distance at formation (see, e.g., Equation \ref{eq:f_GW_peak}), and that $t_{\rm GW} \approx t_{\rm GW}^{e=0} \left(1- e^2 \right)^{7/2}$ \citep{Peters1964}, where $t_{\rm GW}^{e=0}$ refers to the GW merger time for a circular binary, one finds that $f_{\rm{GW}}^{\rm{peak}}$ for a BBH with initial semi-major axis $=a$, BH mass $=M$, and GW merger time $t_{\rm GW} = \tau$ can be expressed as
\begin{multline}
\frac{f_{\rm GW}^{\rm{peak}}}{\text{Hz}} \approx  2 \times 10^{-5} \left(\frac{\tau}{10^{10}\text{yr}}\right)^{-3/7} \\ \times \left(\frac{a}{0.5\text{au}}\right)^{3/14} \left( \frac{M}{30M_{\odot}}\right)^{-11/14}.
\label{eq:fp}
\end{multline}
Here, we have normalized to values that are typical for a BBH near ejection (see Figure \ref{fig:3panel}), as this is where the majority of the relevant dynamics take place. Note that this same relation was derived and implemented in \citep{D'Orazio2018}.

Substituting the time scales, $\tau$, for the three channels into Equation \ref{eq:fp}, one finds that $\log f_{\rm{GW}}^{\rm{peak}}/{\rm Hz} \approx -4.5$ for the ejected mergers ($\tau = 10^{10}$ years), $\log f_{\rm{GW}}^{\rm{peak}}/{\rm Hz} \approx -3.5$ for the in-cluster mergers ($\tau = 10^{7}$ years), and $\log f_{\rm{GW}}^{\rm{peak}}/{\rm Hz} \approx 0$ for the GW capture mergers ($\tau = 0.1$ year). In addition, solving for the corresponding initial eccentricity, $e_{0}$, of the BBHs from each channel, one finds that $e_{0} \approx 1-10^{-1}$ (ejected mergers), $e_{0} \approx 1-10^{-2}$ (in-cluster mergers), and $e_{0} \approx 1-10^{-5}$ (GW capture mergers). From these simple arguments, it follows that in-cluster mergers generally will appear eccentric near the LISA band, and GW capture mergers near the LIGO band, as further discussed in \citep{Samsing2018a,Zevin2018}, and as is clearly shown in Figure \ref{fig:multiband}.
Note that our derived $f_{\rm{GW}}^{\rm{peak}}$ and $e_{0}$ represent minimum values, as our introduced time scales, $\tau$, represent upper limits; in principle all BBH merger channels will have some contribution at high $f_{\rm{GW}}^{\rm{peak}}$ and $e$, but the probability drastically decreases above their characteristic values derived above.

The location of $f_{\rm{GW}}^{\rm{peak}}$ for the in-cluster and GW capture mergers depends especially on the value of the semi-major axis where dynamical ejection is possible, $a_{\rm ej}$, which is about where $a_{\rm ej} \propto M/v_{\rm esc}^2$, where $v_{\rm esc}$ is the escape velocity of the cluster \citep[e.g.][]{Samsing2018}. At this semi-major axis, the time between strong binary-single encounters can be approximated by $1/\Gamma_{\rm enc} \propto v_{\rm esc}^{3}/(n_{\rm s}M^{2})$, where $\Gamma_{\rm enc}$ is the encounter rate, $n_{\rm s}$ is the number density of single BHs, and we have assumed that the ratio between the velocity dispersion and the escape velocity is constant. Correspondingly, the BBH orbital time at $a_{\rm ej}$ is $\propto v_{\rm esc}^{-3}M$, which simply follows from Kepler's law. Now substituting these two time scales into Equation \ref{eq:fp} with $a = a_{\rm ej}$, we find that $f_{\rm{GW}}^{\rm{peak}} \propto v_{\rm esc}^{-12/7}n_{\rm s}^{3/7}M^{2/7}$ for the in-cluster mergers, and $f_{\rm{GW}}^{\rm{peak}} \propto v_{\rm esc}^{6/7}M^{-1}$ for the GW capture mergers.

As seen, these two channels are expected to give rise to $f_{\rm{GW}}^{\rm{peak}}$ positions that scale in very different ways with the BH mass and cluster parameters. In this idealized picture, this interestingly suggests that the host cluster properties might be extracted from future GW data taking into account the actual $f_{\rm{GW}}^{\rm{peak}}$ values for each channel, as well as their relative spacing in $f_{\rm{GW}}^{\rm{peak}}$. This opens up very interesting possibilities if one further considers observatories like DECIGO \citep{2011CQGra..28i4011K, 2018PTEP.2018g3E01I} or Tian Qin \citep{2016CQGra..33c5010L}, as explored in e.g., \citep{Chen2017}. Finally, for a demonstration of how the analytical formalism used above also can be used to accurately explain the relative rate for each of the three channels we refer the reader to \citep{Rodriguez2018b} to estimate the relative rate for each of the three channels.


\section{LISA sources in the local universe}
\label{sec:local_universe}

In order to predict the number of BBHs dynamically-assembled in GCs that will be resolvable by LISA, we must introduce a scheme to build a representative local universe of GCs using our models. Here we use a weighting scheme similar to \citep{Rodriguez2018b} where models are assigned weights based on the present-day GC mass function and metallicity distributions of \citep{El-Badry2018}.

We consider three values for the number density of GCs, $\rho_{\rm{GC}}$, in the local universe \citep{Rodriguez2015a,Rodriguez2016a}: $0.33\,\rm{Mpc}^{-3}$ (pessimistic), $0.77\,\rm{Mpc}^{-3}$ (fiducial), and $2.31\,\rm{Mpc}^{-3}$ (optimistic). For each of the three values of  $\rho_{\rm{GC}}$, we make an appropriate number of draws from our (weighted) grid of 24 GC models. Each drawn cluster model is then assigned a distance (drawn from a uniform volume of radius 500 Mpc) and age (drawn from the metallicity-dependent age distributions of \citep{El-Badry2018}).  The effective time at which a particular cluster would be observed is then given by $t_{\rm{effective}}=t_{\rm{age}}-t_{\rm{lookback}}$, where $t_{\rm{lookback}}$ is the lookback time corresponding to the luminosity distance drawn for each model.

We then look at all BBHs appearing in the particular model (including both BBHs still retained in the host cluster as well as BBHs that were ejected earlier in the cluster evolution, but have not yet merged) in the time window bounded by $t_{\rm{effective}}$ and $t_{\rm{effective}} + T_{\rm{obs}}$, where $T_{\rm{obs}}$ is the duration of the LISA mission. We consider missions of length 2 years, 5 years (fiducial), and 10 years, with the later serving as an upper limit on the observation time.

We perform this procedure an appropriate number of times, based on the particular value of $\rho_{\rm{GC}}$, to generate a realistic population of GC BBHs expected in the local universe. The vast majority of BBHs in this population will be found either within or below the LISA frequency range ($10^{-5} < f_{\rm{GW}} < 1\,$Hz) during the entire LISA observation window. However, a small handful of binaries that happen to be caught when their inspiral time is less than $T_{\rm{obs}}$, will pass through the upper portion of the LISA band and merge in the LIGO band during the LISA observation window.

\begin{figure}
\begin{center}
\includegraphics[width=0.5\textwidth]{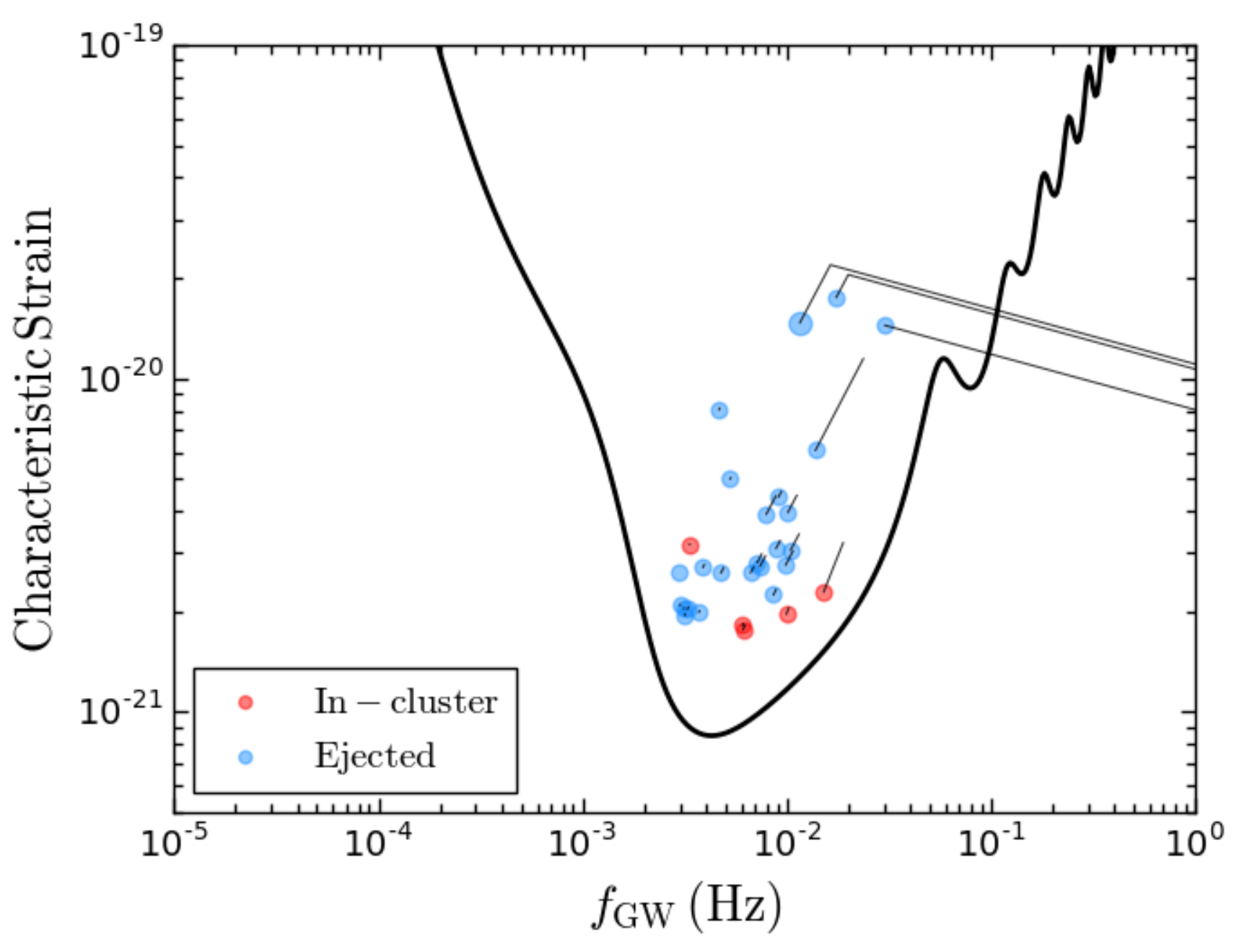}
\caption{\label{fig:LISA} All BBHs formed in GCs in a single local-universe realization that are resolved with $S/N>2$ during a 5-year LISA observation. Blue systems have been ejected from their host cluster, red systems are still retained. Black curves denote the evolution of chirping systems through strain-$f_{\rm{GW}}^{\rm{peak}}$ space during the LISA observation. For our fiducial value of $\rho_{\rm{GC}} = 0.77\,\rm{Mpc}^{-3}$, we predict approximately 25 (2) sources will be resolvable with $S/N > 2\,(5)$, including up to roughly 5 resolvable BBHs ($S/N>2$) that merge in the LIGO band during the LISA observation.}
\end{center}
\end{figure}

For all BBHs identified in the LISA band using the scheme, we calculate the signal-to-noise ratio ($S/N$), based on the binary orbital parameters and the heliocentric distance for each host cluster (randomly drawn as described above) . For eccentric binaries, the $S/N$ is calculating by summing over all relevant harmonics \citep[e.g.,][]{O'Leary2009}:

\begin{equation}
\label{eq:SNR}
\Big(\frac{S}{N} \Big)^2 = \sum_{n=1}^{\infty} \int_{f_{\rm{start}}}^{f_{\rm{end}}} \Big[ \frac{h_{c,\,n}(f_n)}{h_{f}(f_n)} \Big]^2 d \ln{f_n},
\end{equation}
where $h_{c,\,n}$ is given by Equation \ref{eq:hcn}, and $h_f$ is the characteristic LISA noise curve, which we take from \citep{Cornish2018}. $f_{\rm{start}}=nf_{\rm{orb}}$ is the GW frequency emitted at the $n^{\rm{th}}$ harmonic at the start of the LISA observation and $f_{\rm{end}}$ is either the GW frequency at merger or the GW frequency of the $n^{\rm{th}}$ harmonic of the orbital frequency of the binary at the end of the LISA observation time. The characteristic noise (e.g., the black curve in Figures \ref{fig:multiband}
 and \ref{fig:LISA}) can be expressed as

\begin{equation}
h_f(f_n) = \sqrt{\frac{f_n P_n(f_n)}{\mathcal{R}(f_n)}}
\end{equation}where $P_n(f_n)$ is the noise power spectral density of the detector and $\mathcal{R}(f_n)$ is the frequency-dependent signal response function (incorporating sky and polarization averaging), which is given by Equation 8 of \citep{Cornish2018}. Note that in some papers, an extra factor of two is included in Equation \ref{eq:SNR} to take into account the fact that LISA is a two-channel detector. However, for the sensitivity curves of \citep{Cornish2018}, this factor of two is absorbed into the signal-response function, so we do not include it in Equation \ref{eq:SNR}.

In addition to sky and polarization averaging (which are absorbed into the signal response function, $\mathcal{R}$) inclination averaging must also be included, which introduces a factor of 16/5 \citep[see, e.g.,][]{Cornish2018}. We note for clarity that inclination averaging is already taken into account in our Equation \ref{eq:Edot} (see \citep{Peters1963} for further detail on the derivation of this equation), so no additional factor is necessary in Equation \ref{eq:SNR}.


Figure \ref{fig:LISA} shows the characteristic strain of the peak harmonic and peak GW frequency for all BBHs with $S/N > 2$ in a single representative local universe realization assuming our fiducial values of $\rho_{\rm{GC}}=0.77\,\rm{Mpc}^{-3}$ and $T_{\rm{obs}}=5$ years. Large circles denote binaries resolved with $S/N>10$ and, as before, the blue color denotes binaries that have been ejected and red denotes binaries still retained in their host cluster. The circles mark the position of each binary at the start of the LISA observation and the black tracks show the evolution of each system over the duration of the LISA mission. Systems at lower frequencies ($\lesssim 5$ mHz) with no black tracks do not evolve significantly during the LISA observation time.

The ``turn-over'' points in the evolution tracks shown in Figure \ref{fig:LISA} (as well as those in Figure \ref{fig:multiband}) arise because for a finite observation time, the number of cycles in a given frequency bin cannot exceed $fT_{\rm{obs}}$. The location of these turn-over points corresponds to the frequency at which $\dot{f}_n$ is less than $f_n/T_{\rm{obs}}$ for the harmonic of peak GW emission for each binary (as described above).

For the particular local-universe representation shown in Figure \ref{fig:LISA}, we identify 29 BBHs with $S/N > 2$, four of which have $S/N>5$, and one of which has $S/N > 10$. Of the systems with $S/N > 2$,  24 are ejected binaries and 5 are retained. Additionally, 3 of these resolvable binaries merge in the LIGO band during the assumed 5-year LISA mission.

\begingroup
\renewcommand{\arraystretch}{1.7}
\begin{table}
\begin{tabular}{|P{2cm}|P{2cm}|P{1.7cm}|P{1.7cm}|}
\hline
\hline
$\rho_{\rm{GC}}\,(\rm{Mpc}^{-3})$ & 0.33 & 0.77 & 2.31 \\
\hline
Merger rate ($\rm{Gpc}^{-3}\,yr^{-1}$)& $3.3^{+0.8}_{-1.6}$ & $6.6^{+3.0}_{-2.6}$ & $19.8^{+2.6}_{-3.0}$ \\
\hline
\hline
& \multicolumn{2}{c}{$T_{\rm{obs}}=2$ yr} & \\
 \hline
 $S/N > 2$ & $2.9^{+1}_{-2}$ & $6.7^{+2.3}_{-4.7}$ & $28.3^{+8.3}_{-8.3}$ \\
\hline
$S/N > 5$ & $0.1^{+0.3}_{-0.1}$ & $0.3^{+0.6}_{-0.3}$ & $2.68^{+1.68}_{-1.68}$ \\
\hline
 $S/N > 10$ & 0 & $0.04^{+0.1}_{-0.04}$ & $0.1^{+0.3}_{-0.1}$ \\
\hline
& \multicolumn{2}{c}{$T_{\rm{obs}}=5$ yr} & \\
 \hline
 $S/N > 2$ & $10.5^{+1.9}_{-2.4}$ & $24.6^{+4.4}_{-5.6}$ & $73.8^{+8.2}_{-8.8}$ \\
\hline
$S/N > 5$ & $0.6^{+1.7}_{-0.6}$ & $1.4^{+2.6}_{-1.4}$ & $4.8^{+2.2}_{-3.8}$ \\
\hline
 $S/N > 10$ & $0.1^{+0.4}_{-0.1}$ & $0.2^{+0.8}_{-0.2}$ & $0.5^{+0.5}_{-0.5}$ \\
 \hline
 & \multicolumn{2}{c}{$T_{\rm{obs}}=10$ yr} & \\
 \hline
 $S/N > 2$ & $26.3^{+0.7}_{-1}$ & $61.3^{+1.7}_{-2.3}$ & $183.8^{+2.2}_{-1.8}$ \\
\hline
$S/N > 5$ & $39.4^{+0.3}_{-1}$ & $7.3^{+0.7}_{-2.3}$ & $21.8^{+2.2}_{-0.8}$ \\
\hline
 $S/N > 10$ & $0.3^{+0.4}_{-0.3}$ & $0.8^{+1.2}_{-0.8}$ & $2.3^{+0.7}_{-2.3}$ \\
 \hline
\end{tabular}

\caption{\label{table:1} Row 1 shows the three values assumed for globular cluster number density, $\rho_{\rm{GC}}$: 0.33, 0.77, and $2.31\,\rm{Mpc}^{-3}$ , which correspond to the pessimistic, fiducial, and optimistic cases, respectively. Row 2 shows the local-universe merger rate for these three values of $\rho_{\rm{GC}}$. Subsequent rows show the number of LISA sources predicted with $S/N>2$, 5, and 10 for the three $\rho_{\rm{GC}}$ cases assuming a LISA mission of length $T_{\rm{obs}}=2\,$years and $5\,$years (fiducial). We also show the values for $T_{\rm{obs}}=10\,$years, which serves as an upper limit. Note that all values shown here are averages (with error margins showing upper and lower bounds) calculated from all local universe realizations.}
\end{table}
\endgroup

The number of resolvable BBHs in a particular local-universe representation varies stochastically with a number of (randomly drawn) parameters including cluster ages and heliocentric distances. Therefore, we produce 12 total independent realizations to estimate the level of stochastic fluctuations for the number of BBH sources resolvable by LISA and the BBH merger rates.

Table \ref{table:1} shows the predicted merger rate in the local universe (row 2) for the three values of $\rho_{\rm{GC}}$ (row 1). We also show the number of sources predicted to be resolved with $S/N >2$, 5, and 10 for each of these three values of $\rho_{\rm{GC}}$ for three LISA observation times: 2 years, 5 years (fiducial), and for an upper limit of 10 years. All values shown in Table \ref{table:1} are averages (with error margins showing upper and lower bounds) computed from all local universe realizations.

For our fiducial values of $\rho_{\rm{GC}}=0.77\,\rm{Mpc}^{-3}$ and $T_{\rm{obs}}=5\,$yr, we predict approximately 25 BBHs will be resolvable with $S/N>2$, including roughly 2 with $S/N > 5$. If the LISA mission is extended up to a 10-year observation time, these numbers may be extended up to approximately 60 and 7, respectively, as shown in Table \ref{table:1}.

We note that our predicted merger rate for the local universe (6.6, 3.3, and 19.8$\,\rm{Gpc}^{-3}\,\rm{yr}^{-1}$ for the fiducial, pessimistic, and optimistic $\rho_{\rm{GC}}$, respectively) is consistent with merger rates of BBHs in GCs predicted by similar studies \citep{RodriguezLoeb2018,Rodriguez2018b}. For the fiducial value of $\rho_{\rm{GC}}=0.77\,\rm{Mpc}^{-3}$, we find, on average, 1.5 (and up to 5) of the sources with $S/N >2$ will merge in the LIGO band for a 5-year LISA observation. If the mission is extended up to 10 years and assuming the optimistic case of $\rho_{\rm{GC}}=2.31\,\rm{Mpc}^{-3}$, we predict as many as 10 of the LISA-resolvable sources will go on to merge and be detected by LIGO during the LISA mission.

\begin{figure}
\begin{center}
\includegraphics[width=0.5\textwidth]{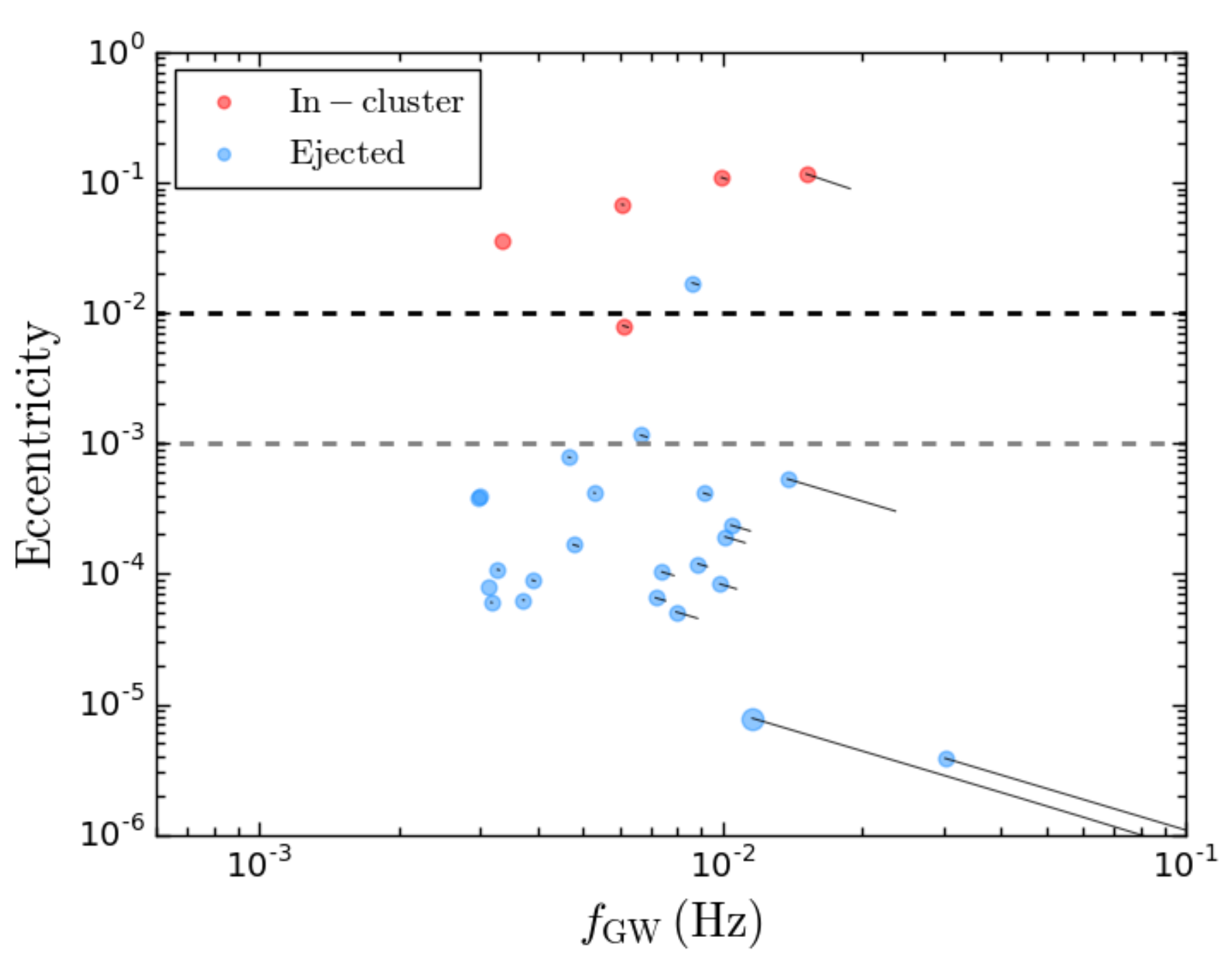}
\caption{\label{fig:ecc_tracks} Evolution of all binaries with $S/N >2$ shown in Figure \ref{fig:LISA} (a single local-universe realization) in eccentricity-$f_{\rm{GW}}^{\rm{peak}}$ space. The upper and lower dashed lines denote the two measurable eccentricity limits as discussed in the text.}
\end{center}
\end{figure}

\subsection{Eccentricity distribution of BBHs in local universe}

One of the exciting prospects for LISA is the potential to measure eccentricities of resolvable sources and use these eccentricities to distinguish between dynamically formed sources, such as those considered here, and binaries that form through other formation channels, in particular, through isolated binary evolution in the Galactic field.

Figure \ref{fig:ecc_tracks} shows the evolution in eccentricity-$f_{\rm{GW}}^{\rm{peak}}$ space of all binaries with $S/N>2$ for the same local universe realization shown in Figure \ref{fig:LISA}. As in Figure \ref{fig:LISA}, the black tracks here show the evolution during the (5 year) LISA observation time. The upper dashed line shows the eccentricity ($e=10^{-2}$) that will always be measurable for any resolved BBH and the lower dashed line denotes the measurable eccentricity limit ($e=10^{-3}$) for $90\%$ of resolvable BBHs for $T_{\rm{obs}}=5$ years \citep{Nishizawa2016}. Note that this figure demonstrates similar results to that Figure 1 of \citep{D'Orazio2018}, as well as Figure 7 of \citep{Banerjee2018}, who found similar eccentricities of in-cluster and ejected BBHs from open-type clusters at similar values of $f_{\rm{GW}}^{\rm{peak}}$.

\begin{figure*}
\begin{center}
\includegraphics[width=\linewidth]{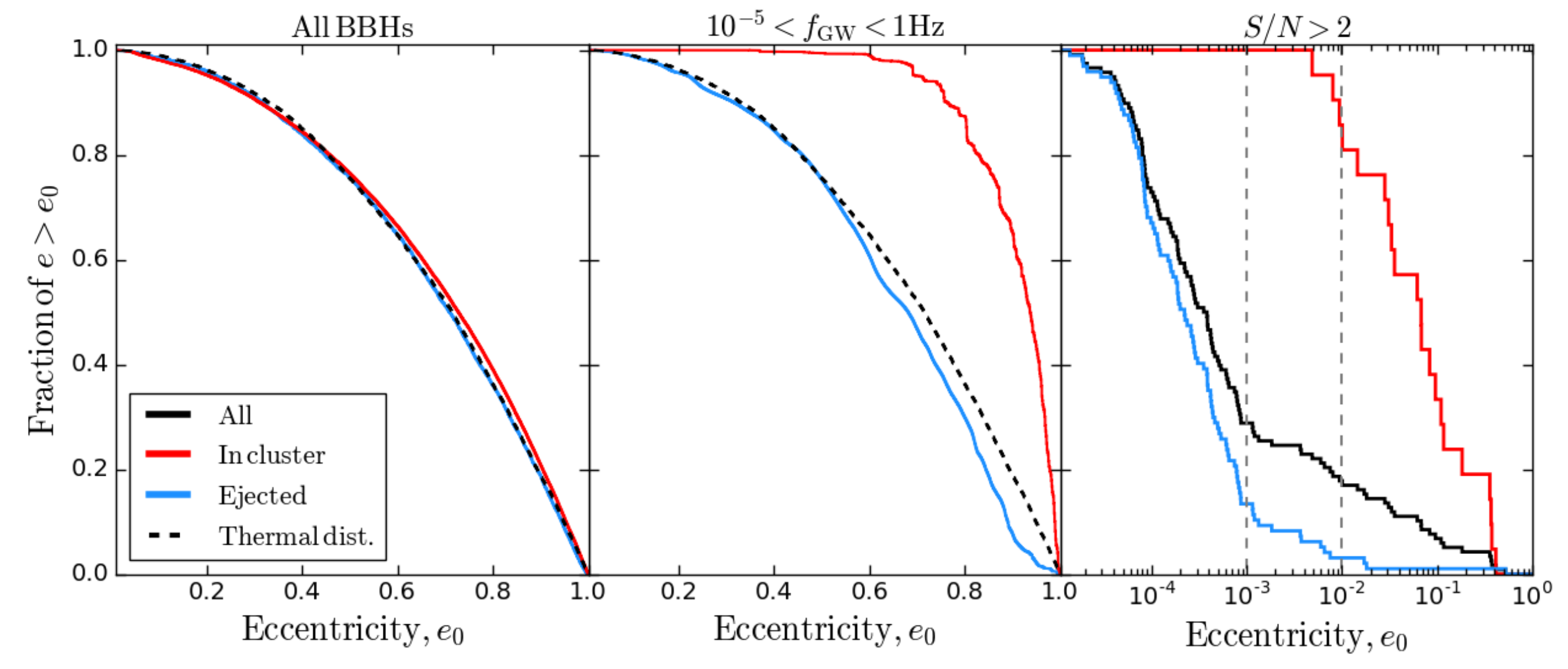}
\caption{\label{fig:ecc_3panel} Eccentricity distribution of all sources found in all computed local universe realizations. Blue denotes binaries that have been ejected from their host cluster, red denotes binaries still retained, and black denotes all binaries. The left panel shows the cumulative distributions for all BBHs of any frequency (the majority of which are found outside LISA band; $f_{\rm{GW}} < 10^{-5}\,$Hz). As expected, this population of binaries conforms to the thermal distribution (dashed black line). The middle panel shows all BBHs found  within the LISA frequency range ($10^{-5} - 1\,\rm{Hz}$). The right panel shows all BBHs resolved with $S/N >2$ in the LISA band. The left and right vertical dashed lines in the right panel show the minimum measurable eccentricity limits discussed in the text ($e=10^{-2}$ and $10^{-3}$).}
\end{center}
\end{figure*}

Figure \ref{fig:ecc_3panel} shows the eccentricity distributions for three populations of BBHs, computed from all local universe realizations considered in this study (as opposed to Figures \ref{fig:LISA} and \ref{fig:ecc_tracks} which show only a single realization). The left panel shows all BBHs of any frequency, the middle panel shows all BBHs in the LISA frequency band, independent of their resolvability, and the right panel shows only those systems that would be resolvable ($S/N>2$). As before, blue and red curves indicate ejected and retained BBHs, respectively, while the solid black curve in the right-hand panel denotes all sources (ejected and retained combined). The vertical dashed lines in the right-hand panel again denote the measurable eccentricity limits for resolved binaries. The black dashed lines in the middle and left-hand panels denote the thermal distribution. 

As the left-hand panel shows, the eccentricity distribution for BBHs of all frequencies is approximately thermal for both ejected and retained BBHs, as is expected for binaries that have undergone dynamical encounters \citep{Jeans1919, Heggie1975,Ambartsu1937,Zevin2018}. Meanwhile, for binaries in the LISA band (middle panel), the distributions begin to diverge from the thermal distribution. The ejected systems (blue curve of the middle panel) are slightly sub-thermal. This is expected because by the time these binaries have inspiraled into the LISA band, the original eccentricities at the time of formation (which would have been drawn from a thermal distribution) will have been partially erased due to circularization effects of GW emission.

On the other hand, as shown by the red curve in the middle panel, retained systems in the LISA band have a super-thermal eccentricity distribution. Unlike the ejected BBHs, that can ``spiral" (and circularize) into the LISA band, the majority of in-cluster BBHs do not have ample time to evolve (until getting significantly altered or ejected by the next close encounter). Hence, the in-cluster BBHs depend mainly on higher eccentricities to be in the LISA band. Indeed, the \textit{orbital} frequency distribution of these in-cluster binaries peaks at frequencies lower than $10^{-5}\,$Hz. Therefore, for wider binaries which will undergo subsequent dynamical encounters in the cluster, we preferentially select the (more abundant) lower orbital frequency systems that have the highest eccentricities, contributing to the super-thermal distribution for retained binaries.

The right-hand panel of Figure \ref{fig:ecc_3panel} shows only BBHs resolved with $S/N>2$. As shown, $84\%$ of retained BBHs (red curve) with $S/N>2$ will have $e>0.01$, while $94\%$ will have $e>10^{-3}$. For ejected BBHs (blue curve), $4\%$ and $14\%$ will have $e$ in excess of the $10^{-2}$ and $10^{-3}$ limits, respectively. 

In total, the ejected binaries dominate the total population of resolvable sources: $83\%$ of BBHs with $S/N>2$ have been ejected from their host cluster. For all BBHs formed in GCs (black curve), we estimate $18\%$ and $30\%$ of sources with $S/N>2$ will have $e>10^{-2}$ and $e>10^{-3}$, respectively. 

Additionally, we note that the eccentricity distributions for the resolvable sources shown in the right-hand panel of Figure \ref{fig:ecc_3panel} agree closely to those shown in the bottom panel of Figure \ref{fig:3panel}, which shows distributions for all BBH mergers integrated back to $f_{\rm{GW}}=10^{-2}\,$Hz. This of course is expected, since most of the resolvable sources have frequencies near this value (Figure\ \ref{fig:LISA}).

\section{Discussion and Conclusions}
\label{sec:discussion}

Using a set of 24 independent GC models developed with our Monte Carlo code \texttt{CMC}, we have explored the population of binary BHs formed in GCs that may be resolved by LISA out to a distance of 500 Mpc. Our key findings include:

\begin{enumerate}
\item{For our fiducial values for number density of GCs in the local universe ($0.77\, \rm{Mpc}^{-3} \rm{yr}^{-1}$) and the duration of the LISA mission ($T_{\rm{obs}}=5\,$years), we estimate approximately 25 BBHs formed in GCs will be resolved in the local universe above mHz frequencies with $S/N>2$, including approximately 2 and up to 4 with $S/N>5$.}
\item{Of these resolvable sources, we predict approximately 2 and up to 5 will inspiral and merge in the LIGO band \textit{during} the LISA observation lifetime.}
\item{If the LISA mission is extended to 10 years, and if we consider a more optimistic value for the GC number density ($2.31\, \rm{Mpc}^{-3}\,\rm{yr}^{-1}$), up to 200 sources may be resolved with $S/N>2$, including in excess of 20 with $S/N>5$. In this case, as many as roughly 10 resolvable sources may merge in the LIGO band during the LISA lifetime.}
\item{Of the GC BH binaries that are resolved, we predict approximately $30\%$ will have eccentricities in excess of $10^{-3}$ and $18\%$ will have eccentricities above $10^{-2}$. Such eccentricities are likely to be measurable by LISA.}
\end{enumerate}


As shown in Figure \ref{fig:3panel}, BBHs formed through isolated binary evolution in the Galactic field have significantly lower eccentricities in the LISA band compared to the dynamical population. Thus, if the eccentricity can be measured for binaries resolved by LISA, it may be used to point to the dynamical origin of these systems. For our predicted resolvable population (Figure \ref{fig:ecc_3panel}), roughly $18\%$ of BBHs have eccentricities in excess of $10^{-2}$. This value is slightly lower than that predicted by \citep{Samsing2018a}, which predicted roughly $40\%$ of BBHs formed in GCs would have $e>10^{-2}$. The difference here arises simply from the relative abundance of ejected versus retained BBHs considered in these two studies. We do note, however, that the eccentricity distributions of our retained and ejected populations individually are very similar to those found in \citep{Samsing2018a} (compare the right-hand panel of our Figure \ref{fig:ecc_3panel} to Figure 4 in \citep{Samsing2018a}). But because \citep{Samsing2018a} considered \textit{all} BBH mergers at a frequency of $10^{-2}\,$Hz, as opposed to only those that would realistically be resolved by LISA, \citep{Samsing2018a} predicted a larger fraction of the population was made up of the retained binaries (which, as shown in Figure \ref{fig:ecc_3panel}, have relatively high eccentricities at $f = 10^{-2}\,$Hz), and thus a larger fraction of the total sources would have eccentricities in excess of $10^{-2}$. When considering only those BBHs that may be resolved by LISA, the fraction of sources with measurable eccentricities likely decreases slightly, as shown here.

Of the BBHs predicted to be resolved with $S/N>2$ (as shown, for example, in Figure \ref{fig:LISA}), we find approximately 80\% have been ejected from their host clusters. However, as shown in \citep{Samsing2018a} (as well as \citep{Rodriguez2018b}), the merger rate of BBHs out to $z \approx 1$ has been shown to be split approximately 50/50 between in-cluster and ejected. There are several reasons LISA may preferentially resolve ejected BBHs. First, because LISA can only resolve binaries out to roughly $0.5\,$Gpc ($z\approx\, $0.1), LISA preferentially sees older clusters when relatively more BHs have been ejected, unlike LIGO which can see younger clusters at higher redshifts. Second, because binaries have relatively higher eccentricities in the LISA band than in the LIGO band, and because retained systems preferentially have higher eccentricities (see Figure \ref{fig:ecc_3panel}), the GW emission of retained sources are more likely to be spread out over many harmonics compared to the ejected binaries. For some parts of parameter space, this may render the binaries less detectable. Furthermore, as shown in the top panel of Figure \ref{fig:3panel}, in-cluster mergers are, in general, wider than the ejected mergers at the time of formation. As a consequence, the in-cluster population are likely to be relatively wide by the time they reach the most sensitive regions of the LISA band and wider binaries are, in general, less visible.
 
In \citep{Samsing2018b}, it was pointed out that inclusion of PN corrections may lead to depletion of stationary sources at high GW frequencies ($\gtrsim 10^{-3}\,$Hz)  in MW GCs, particularly pertaining to \citep{Kremer2018c}, which used Newtonian GC models to show that approximately 5 BBHs may be resolved in the MW GCs and used a rough cut to exclude binaries that may have been handled incorrectly by the Newtonian models.

Although the present study focuses upon higher frequency BBHs en route to merger, we briefly comment here upon the number of lower-frequency BBH sources that may be observed by LISA in the MW, as predicted by our models with PN corrections. Using an identical weighting scheme to that described in Section \ref{sec:local_universe}, but drawing cluster distances from the galactocentric distance distribution of MW GCs, we predict approximately 6 BBHs may be resolved with $S/N > 2$ that formed in MW GCs (including approximately 2 and 1 with $S/N > 5$ and 10, respectively), approximately in line with \citep{Kremer2018c} which predicted 7 and 4 BBHs would be resolved in the MW GCs with $S/N>2$ and 7, respectively. However, as suggested by \citep{Samsing2018b}, we find that inclusion of general relativistic effects depletes the handful of high-frequency MW sources ($f_{\rm{GW}}\gtrsim\,$mHZ) identified in the Newtonian models of \citep{Kremer2018c}. Unlike the binaries considered in Section \ref{sec:local_universe} (which are found at considerably larger distances), these MW sources identified here are found exclusively at low GW frequencies (sub-mHz). This is simply because the timescales on which systems above $\sim$mHz frequencies ``chirp'' through the upper half of the LISA band are relatively small, making them unlikely be seen in the MW, which only contains roughly 150 GCs. Only when we consider a sufficiently large volume, do we access enough GCs to catch a handful of these high-frequency sources. But of course, as we move to a larger volume, the signal-to-noise ratios of these source decrease, which is why we identify no more than $\sim$ dozens of resolvable binaries regardless of assumptions concerning the density of GCs in the local universe.

However, it should be noted that the set of GC models utilized in this study does not necessarily span the full distribution of MW GCs (in particular, missing the core-collapsed clusters which may have few BHs at present; see \citep{Kremer2018d}). A more detailed study of MW clusters is necessary to explore the effects of the inclusion of PN corrections in greater detail.

As illustrated by Figure \ref{fig:multiband}, some BBH mergers that form through GW capture in GCs will form at frequencies above $\sim 1$ Hz, making the prospects for detecting these systems with LISA poor. However, these systems may be observable by proposed decihertz detectors, such as Tian Qin and DECIGO. Thus, future decihertz missions may be necessary to successfully disentangle the formation channel of these binaries \citep[e.g.,][]{Chen2017, Samsing2018c}. We intend to perform a more thorough investigation of the detectability of these BBHs in the decihertz regime in a later study.


\acknowledgments
We thank the anonymous referees for their useful comments and suggestions. This work was supported by NASA ATP Grant NNX14AP92G 
and NSF Grant AST-1716762. K.K. acknowledges support by the National Science Foundation Graduate Research Fellowship Program under Grant No. DGE-1324585. P.A.S. acknowledges support from the Ram{\'o}n y Cajal Programme of the Ministry of Economy, Industry and Competitiveness of Spain, as well as the COST Action
GWverse CA16104. This work was supported by the National Key R\&D Program of China (2016YFA0400702) and the National Science Foundation of China (11721303). M.L.K. and C.S.Y. acknowledge support by the National Science Foundation under grant No.DGE-0948017. S.C. acknowledges support from
CIERA, the National Aeronautics and Space Administration
through a Chandra Award Number TM5-16004X/NAS8-
03060 issued by the Chandra X-ray Observatory Center
(operated by the Smithsonian Astrophysical Observatory for and on behalf of the National Aeronautics
Space Administration under contract NAS8-03060), 
and Hubble Space Telescope Archival research 
grant HST-AR-14555.001-A (from the Space Telescope 
Science Institute, which is operated by the Association of Universities for Research in Astronomy, Incorporated, under NASA contract NAS5-26555. K.B. is grateful to Mads Sorenson for providing the code used to initialize the field population.)


\bibliographystyle{h-physrev.bst}
\bibliography{mybib.bib}


\end{document}